\documentclass[useAMS,usenatbib]{mn2e}
\usepackage{times}
\usepackage{graphicx}
\title[Bimodal abundance pattern in M51: evidence for corotation resonance effects]{Bimodal 
abundance pattern in M51: evidence for corotation resonance effects}
\author[I. A. Acharova et. al.]
{I. A. Acharova, $^{1}$\thanks{E-mail:
achar@phys.rsu.ru (IAA);  
dolly@sao.ru (OAG); 
jacques@iagusp.usp.br (JRDL); 
mishurov@phys.rsu.ru (YNM); 
ntik@sao.ru (NAT)}
O.A.Galazutdinova, $^{2}$
J. R. D. L\'epine, $^{3}$ 
Yu. N. Mishurov,$^{1,2}$ 
\newauthor
and N.A. Tikhonov $^{2}$\\
$^{1}$ Space Research Department, Southern Federal University, 
5 Zorge, Rostov-on-Don, 344090, Russia \\
$^{2}$ Special Astrophysical Observatory of Russian Academy of Sciences, Nizhnij Arkhyz, 
369197, Karachaevo-Cherkessia, Russia\\
$^{3}$ Instituto de Astronomia, Geof\'isica e Ci\^encias 
Atmosf\'ericas, Universidade de S\~ao Paulo, Cidade 
Universit\'aria, S\~ao Paulo, SP, Brazil}

\begin{document}

\date{Accepted 2008 xxxx. Received 2008 xxxx; in original form 2008 xxxx}

\pagerange{\pageref{firstpage}--\pageref{lastpage}} \pubyear{2008}

\maketitle

\label{firstpage}

\begin{abstract}

A chemical evolution model for the bimodal-like abundance distribution in the external galaxy
M51 recently derived on the basis of HST data for more than a half million red supergiants 
is developed. It is shown that, like in our Galaxy, formation of fine structure of 
the radial abundance pattern -- a rather steep gradient in the internal part of the disc 
and a plateau 
in the middle part -- is due to the influence of the spiral arms, the bend in the slope 
of the distribution being arose near the corotation resonance. Our model strongly suggests 
that M51 is surrounded by overabundant gas infalling onto its disc.

\end{abstract}

\begin {keywords}
Galaxies: spiral - galaxies: evolution - galaxies: abundances.
\end{keywords}


\section{Introduction} 

Until recently, it was widely believed that the radial abundance distribution 
(being measured in logarithmic scale) 
along the galactic disc is described by a simple linear function in the 
Galaxy with a constant gradient. This representation is very 
persistent by several researches albeit some irregularities 
in the radial abundance distribution along the galactic disc were pointed out by 
 several researchers of H II regions and planetary nebulae. 
Nevertheless Twarog, Ashman \& Anthony -- Twarog (1997) 
were the firsts who tried to break this oversimplified point 
of view, but only after Andrievsky et al. (2002, see other papers of this series) 
the idea that 
the gradient in the galactic disc is not constant has gained substantial 
foundation. Over a large number of elements the authors showed 
sufficiently definitely that the abundance pattern in the Galaxy 
is a bimodal, i.e., there is a rather steep gradient in the inner  
part of the disc and a plateau (or a shoulder) in the region including 
the Sun, the bending (or fracture) in the slope of the distribution being inside 
the solar circle. 

The results of Andrievsly et al. are quite reliable since they used 
spectroscopic data of Cepheids which, in addition, have 
sufficiently precise distances and due to their brightness are 
seen in a wide spatial region. In our Galaxy, the above abundance 
pattern is also supported by the young planetary nebulae 
(Maciel \& Quireza 1999; Maciel, Costa \& Uchida 2003), OB stars 
(Daflon  \& Cunha 2004), etc.

In a series of papers by Mishurov, L\'epine \& Acharova (2002),  
L\'epine, Acharova, \& Mishurov (2003), 
Acharova, L\'epine \& Mishurov (2005, hereafter ALM), etc., a theory of bimodal abundance 
structure formation was developed. It is based on the idea that galactic density waves, 
responsible for spiral arms, influence the process of enrichment of galactic disk by heavy 
elements, the corotation resonance (where the rotation velocity of galactic matter coincides 
with the velocity of spiral waves) playing a crucial role: the plateau-like distribution forms 
in the vicinity of the disposition of the corotation circle.

The above theory of ALM was applied to our Galaxy. It is well known, here we 
face the problem that owing to solar position in the galactic disc 
and corresponding light extinction, the observational samples do not cover 
the most part of the galactic disc. One should also bear in mind the 
uncertainties in distances to the objects which can sometimes rich tens 
percents. Moreover, we do not know exactly the number of arms in the Galaxy 
and the global structure of its disc. 
On the contrary, in external galaxies we do not face the above 
problem since we see them entirely. Hence, we can hope to construct the 
radial abundance pattern in the whole galactic disc in spite of all uncertainties in 
distances to them. So, it would be very interesting to apply ALM's theory to external spiral 
galaxies.
 
Unfortunately the corresponding observational data were 
very fragmentary. Perhaps the most complete sample of abundances  derived over
HII regions in external galaxies was presented by Zaritsky, Kennicutt \& Hucra 1994. 
In their sample, 
there are several spiral galaxies that demonstrate bimodal 
radial abundance distribution. But as it was shown by Dutil \& Roy (2001) that to draw 
definite conclusions about fine features in  
chemical abundance gradient in a galaxy one needs sufficiently large number
of objects (more than 16 HII regions).

A completely different observational material was analyzed by Tikhonov, Tikhonov \& Galazutdinova 
(2007, hereafter TTG). 
For the galaxy M51 they carried out the stellar photometry of HST images and determined magnitude $I$ and
color index $V-I$ for more than 0.5 million stars.  
It enables to derive the distribution of the color index of red supergiants along the galactic 
radius. 
According to TTG the distribution of the above index along the galactic radius 
demonstrate a bimodal-like structure. Since 
the color index of red supergiants depends on their metallicity the radial distribution of the last value 
will demonstrate the same radial distribution.

The goal of the present paper is to apply the theory of ALM 
to explain the formation of bimodal-like radial abundance pattern 
in the external spiral galaxy M51.

\section{Basic equations. Gaseous and stellar density distributions}

To take into account the effects of spiral arms on the galactic disc enrichment by heavy elements we use equations from ALM:

\begin {equation}
\frac{\partial \mu_s}{\partial t}=(1-R)\psi ,
\end {equation}
\begin {equation}
\frac{\partial \mu_g}{\partial t}=-(1-R)\psi+f  ,
\end {equation}
\begin {equation}
\mu_g \frac{\partial Z}{\partial t}=P_Z\psi+f(Z_f-Z)+
\frac{1}{r}\frac{\partial}{\partial r}(r\mu_gD\frac{\partial Z}{\partial r}) ,
\end {equation}
where $Z$ is the mass content (or abundance) of heavy elements, $\mu_s$ and $\mu_g$ are 
the surface densities for the stellar and gaseous discs correspondingly, $\psi$ is the star 
formation rate (SFR; we use the instantaneous recycling approximation), $R \approx 0.24$ is 
the stellar 
mass fraction returned into interstellar medium (ISM), 
$f$ is the infall rate of matter onto the galactic disc, $Z_f$ is its abundance, $P_Z$ is 
the rate for enrichment of galactic disc by heavy elements (see below), $r$ is the 
galactocentric distance, $t$ is time. The last term in equation (3) describes the heavy 
elements diffusion due to turbulent motions in ISM. Following Mishurov et al. (2002), 
for the diffusion coefficient $D$ we use the gaskinetic estimate,  modeling the turbulent 
ISM by a system of clouds and supposing the values for them close to the ones in our 
Galaxy (see details in ALM).

For the SFR  Schmidt-like approximation was used:

\begin {equation}
\psi=\beta \mu_g^k,
\end {equation}
where $\beta$ is a normalizing factor, the exponent $k$  was adopted $k=1.4$ 
(Schuster et al., 2007, Kennicutt 1998). The coefficient $\beta$ is a constant both in $r$ and 
$t$. It is fitted so as to derive a plausible final (i.e., present) 
density distributions for stellar and gaseous components.

In the classical paper by Tinsley (1980), the coefficient $P_Z$ was considered as a true 
constant that means the stellar mass fraction ejected into ISM (per unit of time and unit 
of surface in galactic plane) as newly synthesized heavy elements. It is obvious, such 
interpretation was based on a tacit assumption that sources of heavy elements are uniformly 
distributed in galactic azimuth. However, following Oort (1974) we believe that 
the  sources are concentrated in spiral arms. Hence  the rate for 
enrichment of the ISM by heavy elements, $P_Z$, is to be proportional to the frequency at 
which any elementary volume of gas enters spiral arms and occurs close to the sources of heavy 
elements, i.e.:

\begin {equation}
P_Z=\eta |\Omega(r)-\Omega_P|\Theta,
\end {equation}
where  $\Omega(r)$ is the angular rotation velocity 
of matter in galactic disk, $\Omega_P$ is the angular rotation velocity for spiral density 
wave pattern, $\eta$ is the constant for the rate of enrichment the galactic disc by heavy 
elements,  $\Theta$ is some cut-off factor 
(see below). 

It is well known that the spiral wave pattern rotates as a 
solid, i.e. $\Omega_P$ is a constant (Lin, Yuan \& Shu 1969).
The location of the corotation resonance ($r_c$) in a galaxy is determined by:

\begin {equation}
\Omega(r_c)=\Omega_P.
\end {equation}

From the above equations it is seen: in the vicinity of the corotation resonance the 
enrichment of ISM by heavy elements is depressed since the galactic matter here moves 
in phase with spiral waves, so, in the corotation vicinity, the most part of interstellar gas   
(that at the initial moment of time was beyond spiral arms) will never enter the arms. Hence, 
the corresponding gas will be far from sources of heavy elements and will not be enriched by 
them. That is why we expect some irregularity in radial distribution of abundance (say, 
bending or fracture) close to the corotation.

Modeling the abundance evolution in the disk of our Galaxy, ALM referred mainly to oxygen. 
The fact is that oxygen is produced by short-lived SNe II whereas they synthesize only about 
30 \% of iron, 70 \% of that being produced by SNe Ia. SNe II are strongly concentrated in  
spiral arms. On the contrary, for a long time it was believed that SN Ia are long-lived objects 
and they do not keep the memory of the fact that they were born in spiral arms. So we can 
expect that oxygen 
is the best indicator of spiral arm influence on abundance distribution in galactic discs. 
This is why ALM concentrated on explanation of oxygen distribution. 

However, new data show that there are 2 populations
 of progenitors for SNe Ia –- short-lived and long-lived (Mannicci, Della Valle \& Panagia 2006; 
Matteucci et al. 2006).
The first type is to be concentrated in spiral 
arms, the second subgroup is not strongly confined to arms (by the way, perhaps these new 
data explain the old result of Bartunov, Tsvetkov \& Filimonova 1994, who revealed that 
SNe Ia also 
demonstrate correlation with spiral arms, not so sharp as SNe II and SN Ib do, but the 
correlation is clear). Each type of SNe Ia produces about equal part of iron (Matteucci et al. 
2006). In total, SN II and 
short-lived group of SNe Ia synthesize $\sim$ 65\% of iron. So, the total (O + Fe + ...) content 
of heavy element, i.e., the value $Z$ in our notation, may be considered as a sufficiently 
good indicator of spiral arms influence on fine radial abundance distribution in galactic 
disc although not so sharp as 
oxygen. Nevertheless the research on nucleosynthesis in galactic disc should be continued 
since a part of heavy elements sources 
does not concentrate in spiral arms (see Acharova, Mishurov \& L\'epine 2008) .

Several words about the cut-off factor $\Theta$. In the density wave theory of Lin et al.
(1969) the wave zone is restricted by inner and outer Lindblad resonances, the locations for the 
resonances being determined by the condition $\nu  = \pm 1$, where 
$\nu(r)=m(\Omega-\Omega_P)/\kappa$ is the dimensionless wave frequency, 
$m$ is the number of arms (for M51 $m=2$), $\kappa$ is the epicyclic frequency 
(see details in Lin et al., 1969). So we adopt: $\Theta = 1$ if $|\nu_m| \leq 1$ 
and $\Theta = 0$ otherwise.

Following Lacey \& Fall (1985) and Portinari \& Chiosi (1999) the infall rate of matter  
onto the galactic disk is: $f(r,t)=f_0 \exp(-r/\Delta - t/\tau)$, where $f_0$ is 
the central rate, $\Delta$ and $\tau$ are the space and time scales 
(in order to reduce the number of free parameters we suppose $\tau$ to be independent of $r$). 

The above system of equations splits into 2 groups -- the first one that describes the 
evolution of stellar and gaseous densities (equations 1, 2 and 4) and the second one 
that describes 
the evolution of abundance (equations 3, 5). The first group of equations represents a system 
of ordinary differential equations 
that can be solved independently. Equation (3) contains SFR function $\psi(r,t)$ and gaseous 
density $\mu_g(r,t)$ that must be computed before solving it. Unlike the previous group 
of equations, the last one is a partial differential equation. Besides the initial 
conditions we have to superimpose the boundary conditions. For that we use the typical 
conditions for the diffusion processes: the absence of diffusion flows at the galactic 
center and at the outer end (the result does not distinctly depends on value of galactic 
radius -- 15 kpc or 25 kpc). They simply guarantee the abundance finiteness here. As an 
initial conditions for densities we adopt $\mu_g = \mu_S = 0$, for content $Z = 0.1 Z_{\odot}$ 
 (according to Asplund, Grevesse \& Sauval 2005 $Z_{\odot} = 0.0165$).

In Fig. 1 are shown the present radial profiles of gaseous, stellar and full densities in M51 
computed for the following set of parameters: 
$\beta = 0.12 \, (M_{\odot} pc^{-2})^{1-k} Gyr^{-1}$, 
$f_0 = 600 \, M_{\odot} pc^{-2} Gyr^{-1}$, 
$\Delta = 3.5 \, kpc$, $\tau = 3 \, Gyr$. 
In all our modelling we suppose that the distance to M51 is 9.6 Mpc and the age 
of the galactic disc is 10 Gyrs.

The final stellar mass in M51 happens to be of the order of $3.39\cdot10^{10} \, M_{\odot}$, 
gaseous $7.7\cdot10^{9} \, M_{\odot}$, present SFR is $3 \, M_{\odot} yr ^{-1}$. 
The  global parameters are close to the observed ones for M51 (Schuster et al. 2007).


\section{Observational data}

\subsection{\it Radial abundance pattern in the disc of M51}

The above theory refers to the radial abundance distribution. That means that the corresponding 
equations are derived from more common equations by means of their averaging over the azimuth.
However, usually the abundances in external galaxies are derived over HII regions which are
strongly concentrated in spiral arms. Hence, HII regions do not represent the abundance 
distribution over the whole galactic disc. 

As we wrote in the Introduction, for M51 TTG used completely different objects: red supergiants
(the corresponding images were obtained by HST). These stars are found both in spiral arms and 
in the interarm regions. So, we can average the abundance over the azimuthal angle in the galactic 
plane. To do this, the galactic plane was divided into circular rings (ellipses for the visible
disc) of width about $1 - 1.5 \, kpc$ and the mean color index $V-I$ was derived
for each ring over stars that have fallen into the corresponding ring.
Further, using isochrones of Bertelli et al. (1994) we transformed the color index into 
abundance and derived the radial distribution of $Z$ in M51. 
The result is shown in Fig. 2. Here we use a 
customary representation for abundance in logarithmic scale normalizing it to the 
solar value $Z_{\odot}$:

\begin {equation}
[Z] = log (Z) -– log (Z_{\odot}).
\end {equation}

First of all we would like to stress that the abundance in M51 is 
higher than the one in our own Galaxy by about $0.2 \, dex$. 
Data of other authors confirm this conclusion.  Indeed, comparing the radial distribution of 
oxygen in M51 (Boissier at al., 2004) and in the disc of our Galaxy (Portinari \& Chiosi 1999) 
one can see that  the content in M51 is systematically higher than the oxygen abundance 
in our Galaxy by about the same value. What is the reason for such different abundance in M51 
and in the Galaxy? Zaritsky 
et al. (1994) connect it with some global properties of a given galaxy. We show below that it 
demands that the gas infalling onto the disc of M51 is overabundant relatively to the gas 
infalling onto our Galaxy.

Another interesting feature is the bending of the slope of distribution at 
$r \approx 5 \, kpc$ clearly seen in this figure. Indeed, inside 
$r \approx 5 \, kpc$ there is a gradient of the order of $ -0.03 \,dex \,kpc^{-1}$. 
Between $r \approx 5 \, kpc$ and 
$8 \, kpc$ there is a plateau (or a shoulder) like in the Galaxy (see the cited papers of 
Andrievsky et al., and ALM). The explanation of this structure is the target of the present 
paper.

It is worth-while to notice, according to our equations, the content $Z$ means the abundance 
of interstellar gas. Hence the theoretical abundance at present time 
must be compared with the one of young objects. In this connection, 
it is important that stars in TTG sample have ages within 40 Myrs. 

\subsection{\it Rotation curve}

For computation of abundance evolution we need the data on the rotation curve in M51. 
 Several rotation curves were derived, e.g., Sofue (1994), Garc\'ia-Burillo, Gu\'elin \& 
Cernicharo
(1993), etc. These curves differ in the middle and outer parts of M51
(see discussion in Schuster et al., 2007). In the middle part, the rotation velocity derived 
by Sofue is  
several tens km/sec larger than the one of Garcia-Burillo et al. 
But in the outer part, the behavior of these curves differs fundamentally:
the curve of Garc\'ia-Burillo et al. is flat whereas the one of Sofue 
declines too fast -- faster than for the 
Keplerian law. Such behavior of the rotation curve is interpreted 
by Sofue as a consequence of the wrap of the galactic disc due to interaction with the 
M51 companion.

In our modeling we treated 2 types of rotation curves. In the inner part of M51 ($r < 7 \,kpc$) 
the curves are close to the one of Sofue (1994) for both the cases. In the outer part 
they differ: in the first case, the curve is flat ($V_{rot}$ is at level 250 km/s), 
in the second case the curve corresponds to opposite limiting 
case -– it follows the Keplerian law. The rotation curves used in the present paper are 
shown in Fig. 3.

\subsection{\it Independent determination of location of corotation resonance }

Since we connect the peculiarities in the radial abundance pattern with the corotation resonance 
it would be useful to have 
some independent indications of a possible location of it. 
In an external galaxy it can be determined by means of several methods: 
{\it i}) kinematic method of Tremaine \&  Weinberg (1984); 
{\it ii}) analyzing the radial distribution of young bright objects;
{\it iii}) studying the radial variation of relative star formation efficiency 
(Cepa \& Beckman, 1989), etc. In our paper, we will refer to the first two methods.

Using a modified Weinberg -- Tremaine method, Zimmer, Rand \& McGraw (2004)
processed the field of velocities of HI in M51 obtained over 21 cm emission and 
derived $\Omega_P = 38 \pm 7 \, km \,s^{-1} \, kpc ^{-1}$. For the adopted
rotation curves this leads to location of the corotation from $r_c \approx 5.5 \,kpc$ 
to about $8 \,kpc$. 

On the other hand, we would like to pay attention to a dip in the radial distribution 
of blue supergiants (i.e., very young objects) 
in M51 at $r \approx 5 \,kpc$ seen in Fig. 8 of TTG. This is a direct indication of the location 
of corotation. Indeed, according to Roberts (1969) galactic spiral shocks (arising when 
interstellar gas flows through spiral arms) are the triggering mechanism of star formation, at 
least of massive bright stars which are strongly concentrated in spiral arms. However, near the 
corotation, the intensity of the shocks becomes small since the relative velocity of 
interstellar gas 
and spiral pattern tends to zero in its vicinity  ($|\Omega - \Omega_P| \to 0$). Hence, near the 
corotation the stimulating effect of spiral arms on star formation will be depressed. So, here 
we have to observe the reduced number of bright stars. Therefore, the dip in the radial 
distribution 
of blue supergiants indeed indicates the location of corotation resonance.

Taking this fact into account, in our modelling, we will consider the location of the corotation 
resonance between $r = 5.5 \, kpc$ and $6.5 \, kpc$.


\section{Modeling of the abundance pattern formation in M51}

Results of our modeling of radial abundance pattern evolution in M51 for the above 2 types 
of rotation curves and 2 locations of corotation resonance are shown in Figs. 4, 5 
(in our experiments, we found $\eta \approx 5.\cdot10^{-5} \, Gyr^{-1}$). 
From these figures it is seen: the radial abundance distribution in the disc of M51 indeed 
reflects the influence of spiral arms. In all cases, a bimodal -- like radial distribution 
of abundance forms with a gradient in the inner part of the galactic disc close to the observed 
one and the plateau in the region $5 -- 8 \, kpc$. The theoretical fine structure of the 
radial abundance pattern is close to the observed one, the bend of the slope of abundance 
distribution being situated in the vicinity of the corotation resonance. 
Hence, like in our own Galaxy, the bend can be considered as an indicator of position 
of corotation in external galaxies.

In Sec 3.1 we noticed that M51 is overabundant relative to our own Galaxy. As a consequence, 
to get good agreement with observations we 
had to adopt the abundance of infalling onto the galactic disc gas to be  
$Z_f \approx 0.023 \,dex$, that is about 7 times higher than usually adopted in modelling of 
our Galaxy (see ALM). 

The question arises: can we explain the observed radial abundance pattern by means of any other  
combination of free parameters, supposing that $Z_f$ is several times smaller than the above 
value, but the rate--constant $\eta$ is larger (correspondingly the rate of heavy elements 
production is higher) and the initial abundance is much higher than the solar abundance? 
The answer is negative: none of our numerous experiments with other 
various combinations of the free parameters gave the abundance pattern which would coincides with  
the observed one. In order to feel the above statement in Fig. 6 we show    
the results of our computations for $Z_f = Z_{\odot}/5$, $Z(t = 0) = 10 Z_{\odot}$ and 
$\eta = 4\cdot 10^{-4} \, Gyr^{-1}$ and $8\cdot 10^{-5} \, Gyr^{-1}$ 
(in the both cases the flat rotation curve 
was used and the corotation resonance was supposed to be situated at $r = 5.5 \, kpc$). From this 
figure it is seen: \\
{\it 1}) high initial value of abundance does not automatically mean that the final
abundance (i.e., the abundance of young objects at present time) will also be high  since heavy 
elements are mainly consumed by low mass stars in preceding times and will never be returned to 
interstellar gas; \\
{\it 2}) since the rate of heavy elements enrichment depends on the difference 
$|\Omega - \Omega_P|$ the increasing of the rate -- constant $\eta$ leads to a very steep 
gradient much steeper than observed.

In other words, we have to recognize that M51 is surrounded by an overabundant gas.

In our simplest theory we do not derive an abrupt decline of the distribution 
in the outer part of the galaxy, beyond 8 kpc, although within errors our model coincides with 
observations. \footnote{Notice that galactic wind takes away all components of ISM and 
does not change the relative abundance of heavy elements, i.e., $Z$.}
One of the way to explain this fact is to adopt that 
the abundance of infalling gas reduces in the outer part of the galactic disk. 
To demonstrate this 
effect we considered the following model: for $r \le 9 \,kpc$ $Z_f = 0.023$, but beyond $9 \, kpc$
$Z_f$ is  
10\% smaller. The result is shown in Fig. 7. From here the reader can see: it is possible to 
explain rapid falling down of galactic abundance in the outskirt of the disc supposing that the 
infalling gas is only slightly depleted there.


\section {Conclusion}

The theory by Acharova, L\'epine \& Mishurov (2005) which explains the formation of bimodal radial 
abundance distribution in a galactic disc under the influence of spiral arms, especially due to 
effects of corotation resonance, was applied to external galaxy M51. As the observational data 
we used the modified results of Tikhonov, Tikhonov \& Galazutdinova (2007) on color index $V-I$ 
of more than 0.5 million red supergigants in M51 derived by means of treatment of HST images. 
To adjust them for our theory we converted the color index into abundance. 

The observed radial abundance distribution in M51 demonstrates the following fine structure: 
i) the bimodal radial distribution with a gradient of the order of 
$-0.03 \, dex \, kpc^{-1}$ in the inner part of the galaxy 
($r < 5 \,kpc$) and the plateau between about $5$ and $8 \, kpc$ 
(hence, there is a bend in the slope of the distribution at $r \approx 5 \, kpc$);
ii) M51 is an overabundant relative to our Galaxy on about $0.2 \, dex$ and more.

Our theory strongly suggests that like in the Galaxy, the formation of the bimodal radial 
abundance pattern  
in M51 is connected with the influence of spiral arms on the process of enrichment the galactic 
disc by heavy elements, the bend in the slope of the distribution being formed in the vicinity 
of the corotation resonance.

Our modelling have led to an unambiguous conclusion: 
it is impossible to simultaneously explain the overabundance in M51 (relative to our Galaxy) and 
the fine structure of the radial abundance distribution by internal properties of the galaxy, 
say, by higher rate for production of heavy elements in the galactic disc and predominantly 
enriched protogalactic gas. We have to suppose that the gas, surrounding M51 and 
infalling onto its disc, is overabundant. How can it be?

After ALM, we would like to attract attention to the problem of derivation of observational 
information on chemical distribution in different objects with different ages, say, HII regions, 
supergigants, planetary nebulae, open clusters in discs of close galaxies. This will enable us to 
derive more refined information on galactic evolution.

\section*{Acknowledgments}

The work was supported in part by the grant of Federal Agency for Education of the Ministry of 
Education and Science of Russian Federation and the grant of Southern Federal University (Russia).


\begin{figure*}
\vspace{174pt}
\includegraphics[width=84mm]{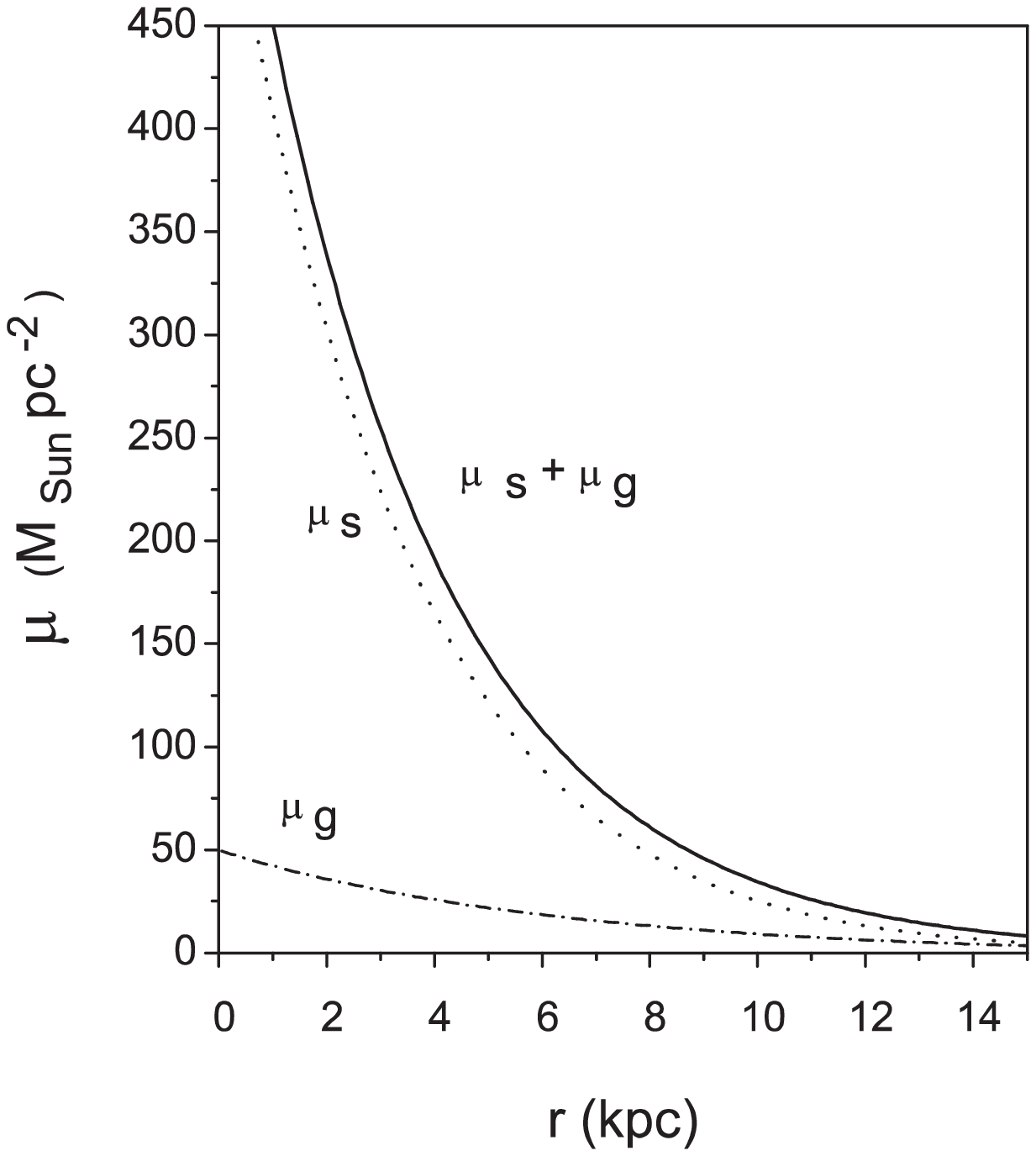}

\caption{Final profiles of surface densities -- gaseous ($\mu_g$), stellar ($\mu_s$) and full 
($\mu_g + \mu_s$) in the disc of M51 at present time. The age of the disc was adopted to be 
10 Gyrs.
\label{f1}}
\end{figure*}

\begin{figure*}
\vspace*{174pt}
\includegraphics[width=84mm]{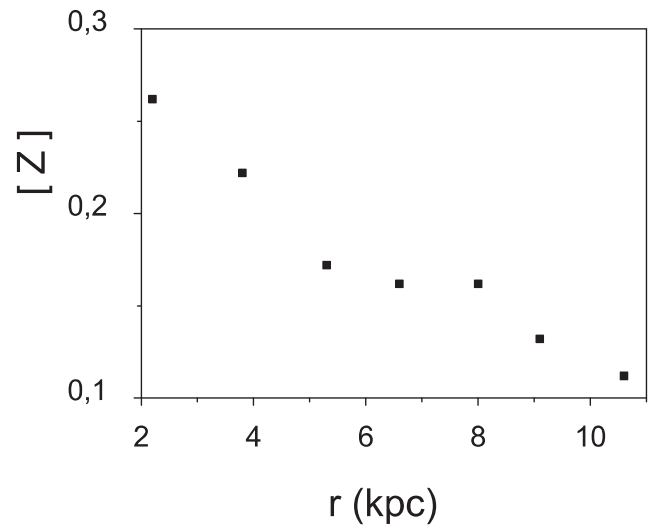}

\caption{Radial distribution of abundance (in logarithmic scale) in M51 normalized on the 
solar abundance (see text). The bimodal structure (in terms of Andrievsky et. al. 2002) with a 
gradient of the order $-0.03 \, dex \, kpc^{-1}$ in the internal part of the disc 
(for $r < 5 \, kpc$) and a plateau between 5 and 8 $kpc$ is clearly seen.
\label{f2}} \end{figure*}


\begin{figure*}
\vspace*{174pt}
\includegraphics[width=84mm]{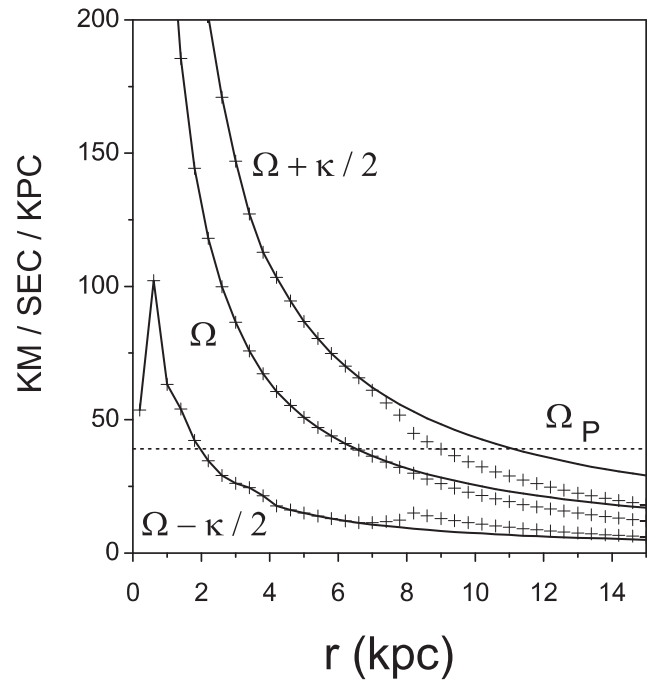}

\caption{Adopted rotation curves: {\it solid line} corresponds to the flat curve in the outer 
part of the disc, {\it crosses} -- to Keplerian law. The {\it dashed line} corresponds to 
$\Omega_P = 39.1 \,km \, sec^{-1} \, kpc ^{-1}$. For this value of the angular rotation velocity 
of spiral wave pattern the corotation resonance happens to be situated at $r = 6.5 \, kpc$. The 
locations of inner and outer lindblad resonances can be derived from intersections of the dashed 
line with the lines corresponding to $\Omega \mp \kappa / 2$.
\label{f3}}
\end{figure*}

\begin{figure*}
\vspace*{174pt}
\includegraphics[width=84mm]{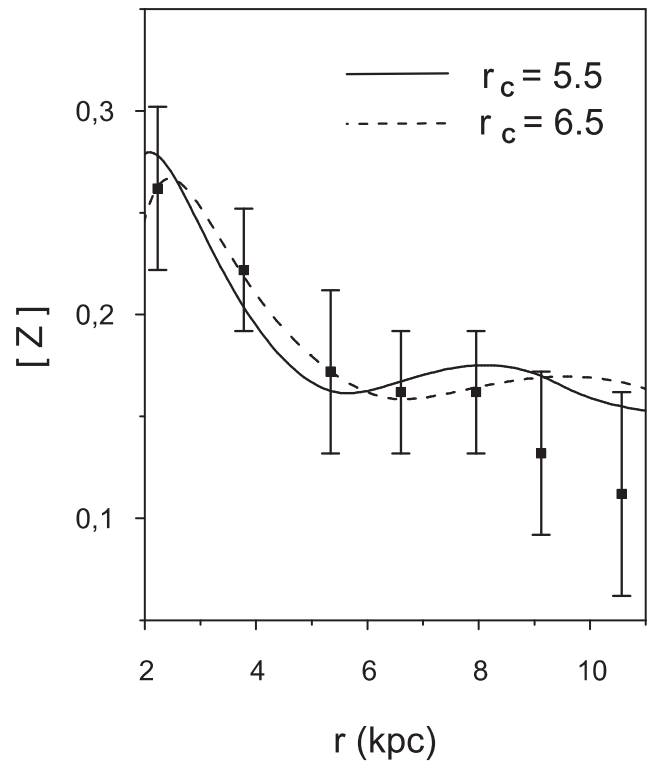}

\caption{Comparison of theoretical radial abundance distribution formed under the influence 
of spiral density waves ({\it solid} and {\it dashed} lines)
for 2 locations of the corotation resonance with the observed pattern  
in the case of flat rotation curve. The observed data are showed by {\it filled squares} 
with error bars.
\label{f4}}
\end{figure*}

\begin{figure*}
\vspace*{174pt}
\includegraphics[width=84mm]{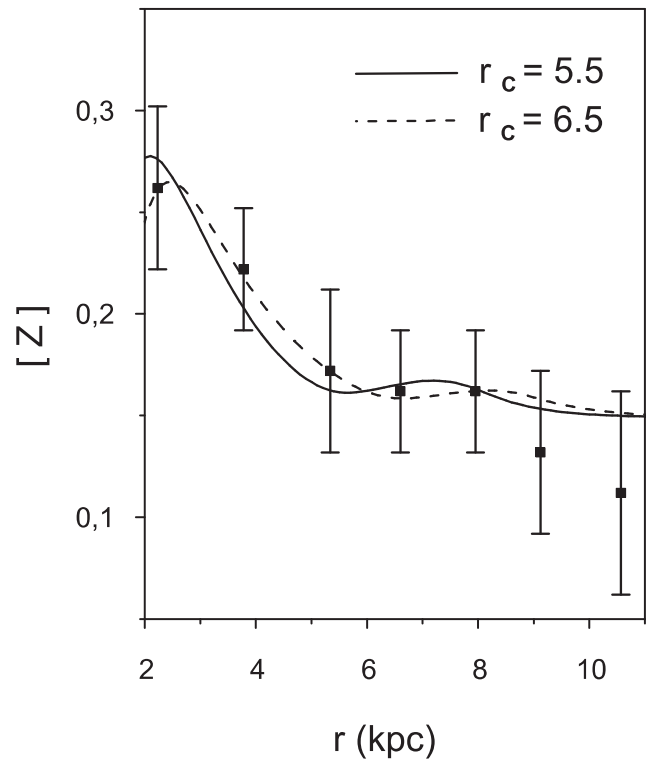}

\caption{The same as in Fig. 4 but for the Keplerian law.
\label{f5}}
\end{figure*}

\begin{figure*}
\vspace*{174pt}
\includegraphics[width=84mm]{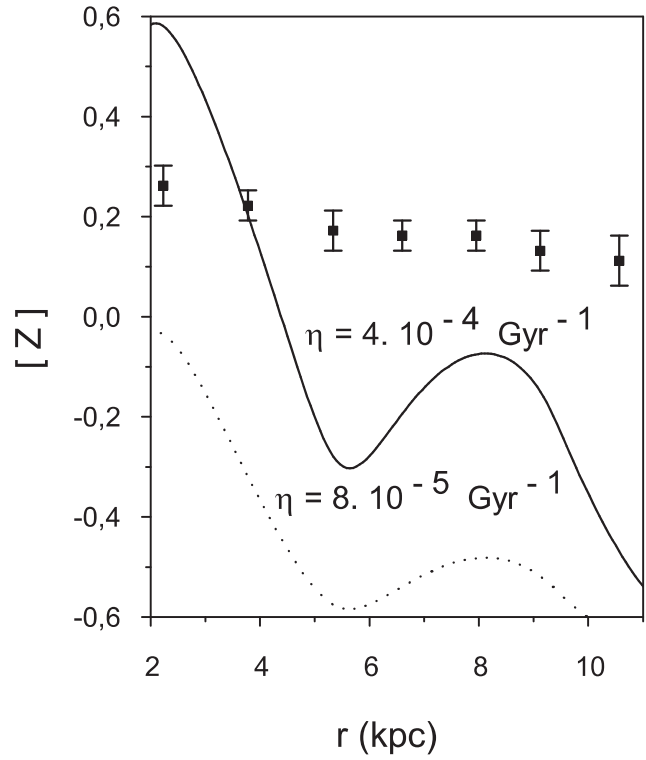}

\caption{The same as in Fig. 4 but for the case when the infalling gas abundance is 
$Z_f = Z_{\odot}/5$, the initial abundance is $Z(t = 0) = 10 Z_{\odot}$, the values of the 
rate--constant 
$\eta$ are shown in the figure and the corotation resonance is situated at $r = 5.5 \, kpc$. 
The observed abundances are given by {\it filled squares} with error bars.
\label{f6}}
\end{figure*}

\begin{figure*}
\vspace*{174pt}
\includegraphics[width=84mm]{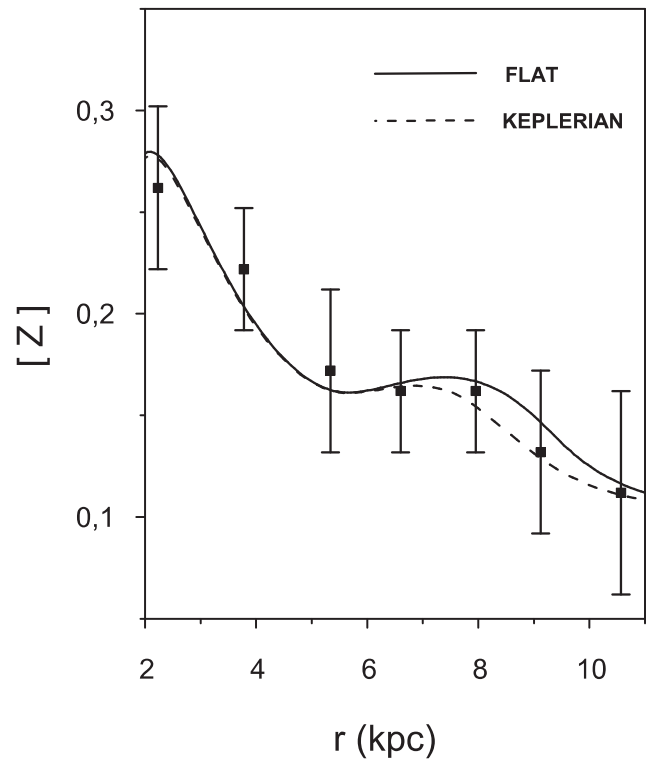}

\caption{The same as in Figs. 4 and 5 but for the case when the infalling gas abundance in 
the outer part of galactic disc ($r>9 \, kpc$) is 10\% less than for $r<9 \, kpc$. For the 
both cases the corotation resonance is situated at $r = 5.5 \, kpc$.
\label{f7}}
\end{figure*}

\label{lastpage}

\end{document}